# Green function of fermions in 2D superconducting Fröhlich model with inhomogeneous order parameter

V. M. Loktev*

*Bogolyubov Institute of Theoretical Physics, National Academy of Sciences of Ukraine, Metrologichna Str. 14-b, Kiev-143, 03143, Ukraine*

V. M. Turkowski

*IIASS ''E. R. Caianiello'', 84019 Vietri sul Mare (SA) and Universitá degli Studi di Salerno, 84081 Baronissi (SA), Italy*

The fermion Green function and spectral characteristics for the 2D Fröhlich model of superconductivity at static fluctuations in the phase of the order parameter are calculated. The results demonstrate strongly non-Fermi-liquid properties of the system at finite temperatures and relate with the pseudogap behavior of high-$T_c$ superconductors at relatively small charge carrier densities. © *2000 American Institute of Physics.*
[S1063-777X(00)00302-9]

## 1. INTRODUCTION

The theoretical description of cuprates with high critical superconducting temperatures remains one of the most exiting and intriguing questions of modern solid-state physics. Because of the electronic and structural complexity of these metaloxide compounds there is a lack of theoretical tools for describing their normal and superconducting properties, which are evidently different from those of low-temperature superconductors.

One of the most interesting peculiarities of cuprates is the presence of a pseudogap in the normal state of samples with lowered carrier densities $n_f$ and temperatures $T$ above the critical value $T_c$. Many theoretical explanations of this phenomenon have been proposed. Among them are explanations based on the model of the nearly antiferromagnetic Fermi liquid,[1] consideration of spin/charge-density waves,[2] and pre-superconducting fluctuations (see, for example, Refs. 3–17). The last characteristic has been studied by many approaches. For example, a $T$-matrix approximation was used in Refs. 3–9. But this approach does not permit a description of ordered states in 2D models (for example, the Berezinskii-Kosterlitz-Thouless (BKT) transition), which are the most suitable for description of cuprates.

It is possible to investigate such states by separating the order parameter (or, for low-dimensional degenerate systems, the so-called complex ordering field) into its modulus $\rho(x)$ and phase $\theta(x)$: $\Phi(x) = \rho(x)\exp[i\theta(x)]$. Although states with $\langle\Phi(x)\rangle \neq 0$ are forbidden in 2D systems at finite temperatures (the Coleman–Mermin–Wagner–Hohenberg (CMWH) theorem,[19]) states with $\rho \equiv \langle\rho(x)\rangle \neq 0$ and $\Phi = \rho\langle\exp[i\theta(x)]\rangle = 0$ can exist.

This approach has been used previously for studying the phase diagram in the $4F$ model[14,15] and in the more realistic Fröhlich model of superconductivity.[16,18] It was shown that in both cases the phase diagram consists of three regions: 1) $T > T_\rho$, where $\rho$ vanishes, i.e., the normal state, 2) $T_\rho > T > T_{BKT}$ ($T_{BKT}$ is the BKT transition temperature), where $\rho \neq 0$ and for $\langle\Phi^*\Phi\rangle$ the correlations decay exponentially, and 3) $T_{BKT} > T$, where these fluctuations have a power-law decay.

This method was also used in Ref. 17 for the one-fermion Green function calculation in the two-dimensional $4F$ model in order to study the fermionic spectral function. It characterises the density of states and allows one to check for the presence of quasiparticle excitations in a system. (For a description of recent experiments on the angle-resolved photoemission spectra (ARPES) of high-$T_c$ superconductors, which contain information about these properties, see Ref. 20. It was shown that right above the critical temperature the width of the quasiparticle peaks noticeably broadens, although the gap in quasiparticle spectrum still remains; this can be connected with the pseudogap properties of the underdoped high-$T_c$ superconductors.

However, the $4F$ model does not take into account many of the properties of real systems, in particular, the retarded nature of the attractive interparticle interaction. As was shown in Refs. 16, and 18, including this property changes the behavior of a system drastically in comparison with the $4F$ case. For example, the width of the region with $T_\rho > T > T_{BKT}$ now goes to zero rather quickly at large charge carrier densities (optimal and overdoped regions).

The aim of this paper is to generalize the results obtained in Ref. 17 to the case of the more realistic Fröhlich model with the retarded interaction. For simplicity we consider the dispersionless ''optical'' phonon mode $\omega(k) = \omega_0 = \text{const}$. Nevertheless, the parameter $\omega_0$ should be considered as the effective weighted value of the frequency of bosons with arbitrary dispersion law $\omega(k)$. This allows us to apply the given approach for an efficient study of any fermion–boson system with fluctuating order parameter.

## 2. THE MODEL

Let us start with the Fröhlich model Hamiltonian density in the standard form:

$$\mathcal{H} = \Psi_\sigma^+(x)\left(-\frac{\nabla^2}{2m} - \mu\right)\Psi_\sigma(x) + g\varphi(x)\Psi_\sigma^+(x)\Psi_\sigma(x) + \mathcal{H}_{\text{ph}}, \quad (1)$$

where $x = \mathbf{r}, \tau$ denotes the space and imaginary time variables; $\Psi_\sigma(x)$ is a fermion field with spin $\sigma = \uparrow, \downarrow$; $m$ is the effective fermion mass; $\mu$ is the chemical potential; $\varphi(x)$ is the phonon field operator, and $g$ is the fermion–phonon coupling constant; we put $\hbar = k_B = 1$. Below we shall use also the Pauli matrices $\tau_1, \tau_2, \tau_3$ in the standard form.

In (1) $\mathcal{H}_{\text{ph}}$ is the Hamiltonian of free phonons with the simplest propagator (in the Matsubara formalism)

$$D(i\Omega_n) = -\frac{\omega_0^2}{\Omega_n^2 + \omega_0^2}, \quad \Omega_n = 2n\pi T, \quad (2)$$

where $\omega_0$, as was pointed out, is the phonon frequency and $n$ is an integer. It was also mentioned in the Introduction that in general this value is the weighted effective frequency of bosons with a momentum-dependent dispersion law $\omega(\mathbf{k})$.

Let us introduce in the Nambu representation $\Psi^+(x) = (\Psi_\uparrow^+(x)\Psi_\downarrow(x))$ the complex superconducting order parameter $\Phi(x) = \Psi^+(x)\tau\Psi(x) = \Psi_\downarrow\Psi_\uparrow$, where $\tau = (\tau_1 - i\tau_2)/2$.

Then in order to study the order-parameter-fluctuation dependences of the Green function

$$G(x) = \langle \Psi(x)\Psi^+(0)\rangle, \quad (3)$$

it is convenient to use the parametrization

$$\Phi(x) = \rho(x)\exp[i\theta(x)]^{1)} \quad (4)$$

with the simultaneous spinor substitution[1)]

$$\Psi(x) = \exp[i\tau_3\theta(x)/2]y(x),$$
$$\Psi^+(x) = Y^+(x)\exp[-i\tau_3\theta(x)/2]. \quad (5)$$

As we have said, we shall consider the situation when $\rho$ is a spatially homogeneous, or constant, quantity and the phase $\theta(x)$ is a random quantity. In fact, the spinors $Y(x)$ and $Y^+(x)$ are none other than the neutral fermion operators. In this case the Green function can be naturally separated into the charge and spin parts (see also, Ref. 17). Namely, in the momentum representation:

$$G(i\omega_n, \mathbf{k}) = T\sum_{m=-\infty}^{\infty}\int\frac{d^2p}{(2\pi)^2}$$
$$\times \sum_{\alpha,\beta=\pm} P_\alpha \mathcal{G}(i\omega_m, \mathbf{p}) p_\beta D_{\alpha\beta}$$
$$\times (i\omega_n - i\omega_m, \mathbf{k} - \mathbf{p}). \quad (6)$$

Here $\mathcal{G}(i\omega_m, p)$ is the Green function of neutral fermions (see, for example, Ref. 16),

$$\mathcal{G}(i\omega_n, \mathbf{k}) = -\frac{i\omega_n \hat{I} + \tau_3 \xi(\mathbf{k}) - \tau_1 \rho}{\omega_n^2 + \xi^2(\mathbf{k}) + \rho^2} \quad (7)$$

with $\xi(\mathbf{k}) = \mathbf{k}^2/2m - \mu$; $D_{\alpha\beta}$ is the correlation function of the phase fluctuations

$$D_{\alpha\beta}(i\Omega_n, \mathbf{q}) = \int_0^{1/T} d\tau \int d^2r \exp[i\Omega_n\tau - i\mathbf{q}\mathbf{r}]$$
$$\times \langle \exp[i\alpha\theta(\tau,\mathbf{r})/2]\exp[-i\beta\theta(0)/2]\rangle \quad (8)$$

and $P_\pm = 1/2(\hat{I} \pm \tau_3)$ are the projectors. The Green function (7) of the neutral fermions coincides identically with that obtained in the $4F$ model, and the electron–phonon (boson) interaction enters this expression through $\rho$, which goes to zero if the coupling constant $g$ (see (1)) vanishes.

It is important to stress once again that in (7) $\rho = $const, i.e., homogeneous. But, of course, the neutral $\rho$ does not play the role of a genuine order parameter in a system, so there is not any contradiction with the CMWH theorem.

## 3. THE GREEN FUNCTION

According to the previous section, for calculation of the Green function (6) it is necessary to know the phase fluctuation correlator $D_{\alpha\beta}$ (8). This quantity can be calculated using a functional integral[17]

$$D_{\alpha\beta} = \int \mathcal{D}\theta(x)\exp\left\{-\int_0^{1/T} d\tau_1 \int d^2r_1\right.$$
$$\left.\times\left[\frac{1}{2}\theta(x_1)D_\theta^{-1}(x_1)\theta(x_1) + I(x_1)\theta(x_1)\right]\right\}$$
$$= \exp\left[\frac{1}{2}\int_0^{1/T}d\tau_1\int_0^{1/T}d\tau_2\int d^2r_1\right.$$
$$\left.\times \int d^2r_2 I(\tau_1,\mathbf{r}_1)D_\theta(\tau_1-\tau_2,\mathbf{r}_1-\mathbf{r}_2)I(\tau_2,\mathbf{r}_2)\right], \quad (9)$$

with the corresponding Green function

$$D_\theta(x) = \langle\theta(x)\theta(0)\rangle \quad (10)$$

of the phase fluctuations and the source of the $\theta$ field

$$I(x_1) = -i\frac{\alpha}{2}\delta(\tau_1-\tau)\delta(\mathbf{r}_1-\mathbf{r}) + i\frac{\beta}{2}\delta(\tau_1)\delta(\mathbf{r}_1).$$
$$\alpha, \beta = \pm. \quad (11)$$

In the second-derivative order the Green function (10) has the form:

$$D_\theta^{-1}(x) = -J(\mu,T,\rho)\nabla_r^2 - K(\mu,T,\rho)(\partial_\tau)^2. \quad (12)$$

The coefficients $J(\mu,T,\rho)$ and $K(\mu,T,\rho)$ have the physical sense of the superfluid stiffness and compressibility, respectively. One can readily obtain the following expressions for them (see, for example, Refs. 16 and 18):

$$J(\mu,T,\rho) = \frac{1}{8\pi}\{\sqrt{\mu^2+\rho^2}+\mu$$

$$+ 2T\ln[1+\exp(-\sqrt{\mu^2+\rho^2}/T)]\}$$

$$-\frac{T}{4\pi}\left[1-\frac{\rho^2}{4T^2}\frac{\partial}{\partial(\rho^2/4T^2)}\right]$$

$$\times \int_{-\mu/2T}^{\infty} dx \frac{x+(\mu/2T)}{\cosh^2\sqrt{x^2+\rho^2/4T^2}},$$

and

$$K(\mu,T,\rho) = \frac{m}{8\pi}\left(1+\frac{\mu}{\sqrt{\mu^2+\rho^2}}\tanh\frac{\sqrt{\mu^2+\rho^2}}{2T}\right.$$

$$-\frac{1}{8}\frac{\rho^2}{4T^2}\frac{\partial}{\partial(\rho^2/4T^2)}$$

$$\left.\times \int_{-\mu/2T}^{\infty} dx \frac{\tanh\sqrt{x^2+\rho^2/4T^2}}{\sqrt{x^2+\rho^2/4T^2}}\right).$$

Note that in comparison with the $4F$ case the functions $J$ and $K$ contain new terms with the derivative. But formally the general expressions for $D_\theta(x)$ in both cases are the same, so let us use below the formulas obtained for the $4F$ model in.[17]

Thus, in the static case $\tau=0$ at $T<T_{BKT}$ and when the coherence length is larger than the lattice spacing (as is justified for cuprates) the correlator has the usual (i.e., power law) form

$$D(\mathbf{r}) = \left(\frac{r}{r_0}\right)^{-T/8\pi J} \quad (13)$$

(this is the expression for the only nonzero components $D_{++}(\mathbf{r},0)$ and $D_{--}(\mathbf{r},0)$). In (13) the quantity $r_0$ is

$$r_0 = \frac{2}{T}\left(\frac{J}{K}\right)^{1/2}. \quad (14)$$

Note that in Ref. 17 it was assumed that $J\sim\varepsilon_F$ (the Fermi energy) and $K\sim$const. Under these assumptions $r_0$ is equal to $2\sqrt{\varepsilon_F/m}/T$ and has the meaning of the single-particle de Broglie wavelength. Whereas these approximations for $K$ and $J$ are justified for the physical regions in $4F$ model, in the boson-exchange case at large carrier densities the asymptotic behavior of $J$ is different ($J\sim$const, so in this region $r_0$ does not have such a simple physical interpretation).

At $T>T_{BKT}$ it was proposed to use for $D(r)$ the expression from the theory of the BKT transition.[21,22]

$$D(\mathbf{r}) = \left(\frac{r}{r_0}\right)^{-T/8\pi J} \exp\left(-\frac{r}{\xi_+(T)}\right), \quad (15)$$

where

$$\xi_+(T) = C \exp\left(\frac{T_\rho - T}{T - T_{BKT}}\right)^{1/2}. \quad (16)$$

This expression could be considered as a general form of $D(\mathbf{r})$ at any temperature if one puts $\xi_+(T)=\infty$ for $T<T_{BKT}$. The constant $C$ can be estimated as $r_0/4$, the value obtained from the assumption that $\xi_+$ cannot be much less than the only natural cutoff $r_0$ in the theory.

Then substitution of expression (7) for $\mathcal{G}$ and the Fourier transform of expression (15) for $D(\mathbf{r})$ into formula (6) results in the next representation for the Green function:

$$G(i\omega_n,\mathbf{k}) = -\frac{Am\xi_+^{2\alpha}}{2\pi\alpha}$$

$$\times\left[\frac{A_1}{(u_1 u_2)^\alpha}F_1\left(\alpha,\alpha,\alpha;\alpha+1;\frac{u_1-1}{u_1},\frac{u_2-1}{u_2}\right)\right.$$

$$\left.+(\sqrt{\omega_n^2+\rho^2}\to-\sqrt{\omega_n^2+\rho^2})\right], \quad (17)$$

where

$$A = \frac{4\pi\Gamma(\alpha)}{\Gamma(1-\alpha)}\left(\frac{2}{r_0}\right)^{2(\alpha-1)}, \quad \alpha = 1 - \frac{T}{16\pi J},$$

$$A_1 = \frac{1}{2}\left(\tau_3 - \frac{\omega_n}{\sqrt{\omega_n^2+\rho^2}}\right), \quad (18)$$

and $F_1$ is the Appell function.[23] The quantities $u_1$ and $u_2$ are defined by

$$u_1 = m\xi_+^2\left(\frac{k^2\xi_+^2+1}{2m\xi_+^2} - \mu + i\sqrt{\omega_n^2+\rho^2} + \sqrt{D}\right),$$

$$u_2 = m\xi_+^2\left(\frac{k^2\xi_+^2+1}{2m\xi_+^2} - \mu + i\sqrt{\omega_n^2+\rho^2} - \sqrt{D}\right) \quad (19)$$

with

$$D \equiv \left(\frac{k^2\xi_+^2+1}{2m\xi_+^2} - \mu + i\sqrt{\omega_n^2+\rho^2}\right)^2$$

$$+ \frac{2}{m\xi_+^2}(\mu - i\sqrt{\omega_n^2+\rho^2}). \quad (20)$$

For studying the spectral properties of the system in the next Section we'll need the retarded Green's function, which can be obtained from (17) after the analytical continuation $i\omega_n\to\omega+i0$.

For now let us just say that for $T<T_{BKT}$ this function has the structure

$$G(\omega,\mathbf{k})\sim\Gamma^2(\alpha)\left(\frac{2}{mr_0^2}\right)^{\alpha-1}A_1[-(\mu+\sqrt{\omega^2-\rho^2})]^{-\alpha}$$

$$\times\left[\frac{\Gamma(1-2\alpha)}{\Gamma^2(1-\alpha)} + \frac{\Gamma(2\alpha-1)}{\Gamma^2(1-\alpha)}\right.$$

$$\left.\times\frac{1}{(1-z_1)^{2\alpha-1}}\right], \quad z_1 = 1.$$

Thus the Green function is of the non-Fermi-liquid theory type; it has a non-pole character and contains a branch cut. So the Fermi-liquid behavior of the system is broken by strong phase fluctuations of the complex ordering field.

## 4. THE SPECTRAL DENSITY AND DENSITY OF STATES

The spectral density contains information about many properties of systems, for examples, such features as the density of states and the presence of a gap. For cuprates this quantity was measured in the ARPES experiments (see Ref. 20). Below we obtain the expressions for the spectral density and density of states which follow from the retarded Green function (recall that it is defined by (17) with the analytical continuation $i\omega_n \to \omega + i0$).

Let us first calculate the spectral density[19] using the expression

$$A(\omega,\mathbf{k}) = -\frac{1}{k}\operatorname{Im} G_{11}(\omega+i0,\mathbf{k}). \tag{21}$$

After substitution of the analytically continued expression (17) in (21) one can directly come to:[17]

$$A(\omega,\mathbf{k}) = \frac{\Gamma(\alpha)}{\Gamma(1-\alpha)}\left(\frac{2}{mr_0^2}\right)^{\alpha-1}\operatorname{sgn}\omega\,\theta(\omega^2-\rho^2)$$

$$\times \left[ \frac{(A_1)_{11}}{D_2^{\alpha/2}} F_1\left(\frac{\alpha}{2}, \frac{1-\alpha}{2}; 1; \right.\right.$$

$$\left.-4\frac{\frac{k^2}{2m}(\mu+\sqrt{\omega^2-\rho^2})}{D}\right)\theta(\mu+\sqrt{\omega^2-\rho^2})$$

$$\left.-(\sqrt{\omega^2-\rho^2}\to -\sqrt{\omega^2-\rho^2})\right]. \tag{22}$$

The chemical potential $\mu$ is determined by the equation that fixes the carrier density.[16] However, in the case of large carrier densities the equality $\mu=\varepsilon_F$ is almost exactly fulfilled. Note that the expression (22) for $A(\omega,\mathbf{k})$ is not the BCS sum of two parts with $\delta$- function peaks at $\omega=\pm E(\mathbf{k})$ which correspond to the addition and removal of an electron, but the sum of two "mixed" terms.

It is possible to check analytically the sum rule for the spectral density. Namely, as in the $4F$ model, we have

$$\int_{-\infty}^{\infty} d\omega A(\omega,\mathbf{k}) = \frac{\Gamma(\alpha)}{\Gamma(2-\alpha)}. \tag{23}$$

Let us estimate the quantity on the right side in the region $T \sim T_{BKT}$. For the stiffness at $T=T_{BKT}$ we have $J=2/\pi T_{BKT}$, which gives [see (18)] $\alpha=1-1/32\simeq 1$ at $T \sim T_{BKT}$. Therefore the formula (22) for the spectral density is quite good in the temperature region-near $T_{BKT}$ at large carrier densities. Since we are studying the region of large carrier densities, at temperatures in the pseudogap phase the condition $T \sim T_{BKT}$ is always true, because at large $n_f$ the pseudogap region is narrow and shrinks as $n_f \to \infty$ (again, see

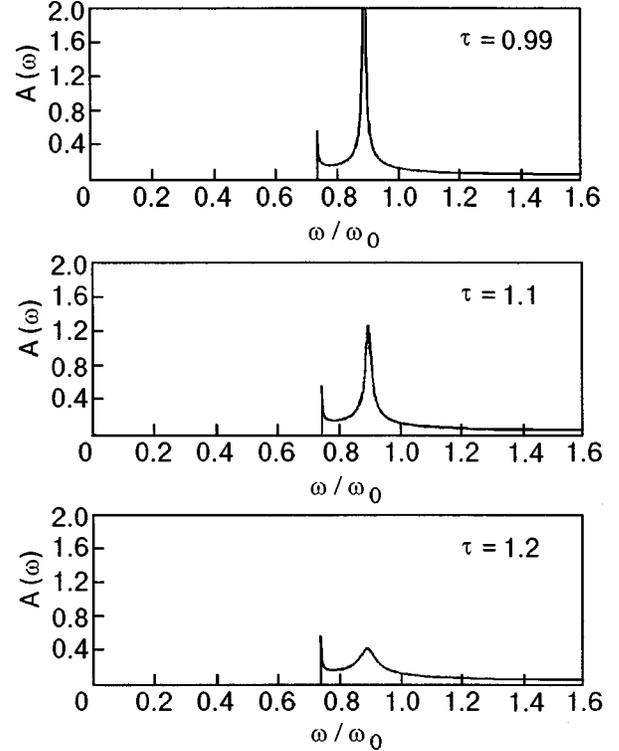

FIG. 1. The spectral density as a function of $\omega$ is presented for the case $k>k_F$ at different values of $\tau=T/T_{BKT}$.

Ref. 16). This is evidently different from the $4F$ case, where the corresponding region decreases much more slowly.

The $\omega$ dependences of the spectral density for $T<T_{BKT}$ and $T>T_{BKT}$ in the case $k<k_F$ are presented in Fig. 1 (the behavior in the case $k>k_F$ is analogous). There are two quasiparticle peaks at the points $\omega=\pm E(\mathbf{k})$ and another two at $\omega=\pm\rho$. The presence of the last two is caused by the non-pole structure of the Green function. At $k=k_F$ these two kinds of peaks coincide, because at this point one has $E(\mathbf{k}_F)=\rho$. The peaks at the frequencies $\omega=\pm E(\mathbf{k})$ decrease with increasing temperature, and when $T>T_{BKT}$ (where these peaks are finite) quickly go to zero. This is in qualitative agreement with the ARPES experiments,[20] which show that the spectral function broadens on passing to the normal phase.

For $\omega<|\rho|$ we have $A(\omega,\mathbf{k})=0$, and therefore the gap exists at any $T$. The same conclusion is also correct for the $4F$ model. Note again that our results are obtained by using the static approximation. The empty region must disappear (as well as the quasipeaks at $\omega=\pm\rho$) if dynamical fluctuations are taken into account. Evidently, the filling of the empty region should be different for $T<T_{BKT}$ and $T>T_{BKT}$.

As is seen in Fig. 1, a smooth crossover takes place as the temperature changes from $T<T_{BKT}$ to $T>T_{BKT}$. This is in agreement with experiments (for instance, on $\rho$)[20] and differs from the BCS theory. Let us also note that our results are obtained for not very small $\rho$. When $\rho \to 0$ (low carrier densities) its (i.e., modulus) fluctuations must be taken into account.

The end of this Section is devoted to a calculation of the density of states. The desired expression can be obtained from the formula

$$N(\omega) = N_0 \int_0^W d\frac{k^2}{2m} A(\omega, \mathbf{k}), \qquad (24)$$

where $N_0 \equiv m/2\pi$ is the density of states in the normal phase, and $W$ is the bandwidth.

This expression together with (22) results in the representation

$$N(\omega) = N_0 \frac{\Gamma(\alpha)}{\Gamma(2-\alpha)} \left(\frac{2}{mr_0^2}\right)^{\alpha-1} \mathrm{sgn}\,\omega\, \theta(\omega^2 - \rho^2)$$

$$\times \Bigg\{ (A_1)_{11} \bigg[\bigg(\frac{1}{2m\xi_+^2} + W - \mu - \sqrt{\omega^2 - \rho^2}\bigg)^{1-\alpha}$$

$$- \bigg(\frac{1}{2m\xi_+^2}\bigg)^{1-\alpha} \bigg] \theta(\mu + \sqrt{\omega^2 - \rho^2})$$

$$- (\sqrt{\omega^2 - \rho^2} \to -\sqrt{\omega^2 - \rho^2}) \Bigg\}, \qquad (25)$$

which formally also coincides with formulas obtained in Ref. 17, although there is different behavior on account of the different carrier dependences of the neutral order parameter $\rho$ in these (4F and boson-exchange) superconducting models.

At zero temperature and large carrier densities ($\mu \gg \rho$) the formula (25) reproduces the BCS result

$$N(\omega) = N_0 \frac{|\omega|}{\sqrt{\omega^2 - \rho^2}}. \qquad (26)$$

The density of states for different cases are presented in Fig. 2. As in the case of the spectral densities, the gap in the density of states exists at temperatures near and above $T_{\mathrm{BKT}}$, which plays the role of the critical temperature in a pure 2D metal. The form of the density of states qualitatively coincides with the BCS one. The crucial difference is in the smooth change of the curves at the phase transition point.

Let us repeat again that dynamical fluctuations can be responsible for the filling of gap and that at small carrier densities the $\rho$ fluctuations must also be taken into account.

## 5. CONCLUSION

In this paper the analytical calculation of the fermion Green function has been generalized to the case of the Fröhlich model of superconductivity, although some expressions have proved to be similar to those obtained for the case of a 2D metal with a nonretarded inter-fermion attractive interaction. This result could be important for several reasons: First, as a general result for the theory of fluctuations in boson-exchange quantum solid state systems. Second, because there is as yet no generally accepted theory of high-temperature superconductivity, and it now appears possible that some boson-exchange model will be appropriate for the description of this phenomenon. Thus an analytical investigation of the Green function in the boson-exchange case could be very important, because it gives much more information than the numerical studies often employed. For example, it has been shown that the transverse phase fluctuations result in non-Fermi-liquid behavior of the system below (but for $T \neq 0$) and above $T_{\mathrm{BKT}}$.

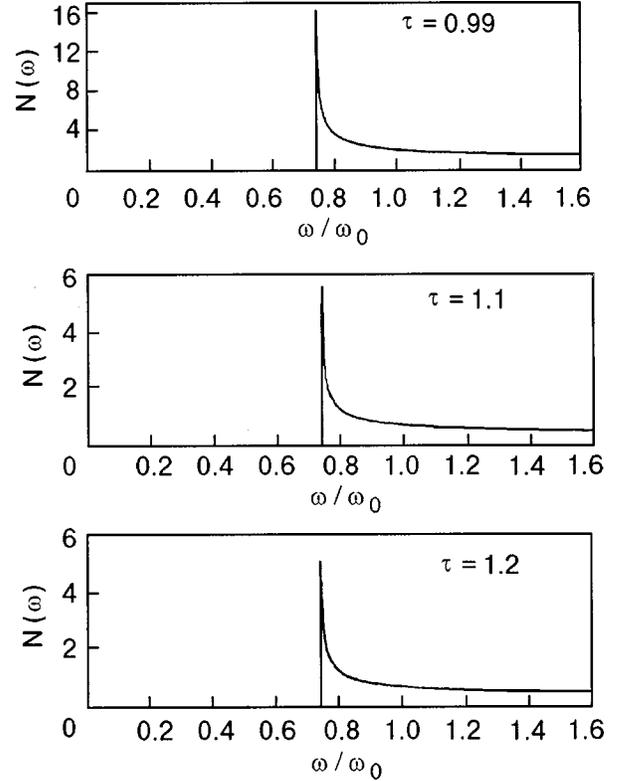

FIG. 2. The density of states (in $N_0$ units) is plotted versus $\omega$ for different values of $\tau = T/T_{\mathrm{BKT}}$.

Along with this there are many open questions about the problem studied above. For example, the role of superconducting fluctuations in the pseudogap phase formation. Also it is very important to take into account the $\rho$ fluctuations and to generalize the approach to the dynamical fluctuation case.

One of us (V. M. T.) acknowledges the financial support of the World Laboratory and also thanks the members of the Dipartimento di Scienze Fisiche ''E. R. Caianiello'' Universita' di Salerno and the International Institute for Advanced Scientific Studies ''E. R. Caianiello,'' Vietri sul Mare (SA), Italy, especially The President Prof. M. Marinaro and Prof. F. Mancini, for hospitality.

*E-mail: vloktev@bitp.kiev.ua
[1)]Here and below we consider $\Psi(x)$ and $Y(x)$ as Grassmann operators and $\theta(x)$ as ordinary variables in functional integrals. In the Hamiltonian formalism the former obey Fermi statistics, while $\theta(x)$ preserves ordinary (commutational) algebra.